\documentclass[12pt]{elsarticle}
\usepackage[utf8]{inputenc}
\usepackage{graphicx}
\usepackage{subfigure}
\usepackage{array}
\usepackage{verbatim}

\usepackage{amssymb}
\usepackage{amsthm}
\usepackage{url}
\usepackage{amsmath}
\usepackage{subfigure}
\usepackage{textcomp}
\usepackage{mhchem}
\usepackage{xcolor}
\graphicspath{{figs/}}
\usepackage[margin=2.7cm]{geometry}

\journal{Good Journal}

\begin{document}

\begin{frontmatter}

\title{On complexity of colloid cellular automata}
\author[a,d]{Andrew Adamatzky*} 
\ead{Andrew.Adamatzky@uwe.ac.uk}
\author[b]{Nic Roberts}
\author[a]{Raphael Fortulan}
\author[a]{Noushin Raeisi Kheirabadi}
\author[a]{Panagiotis Mougkogiannis}
\author[a]{Michail-Antisthenis Tsompanas}
\author[c]{Genaro J. Mart{\'i}nez}
\author[d,a]{Georgios Ch. Sirakoulis}
\author[e,a]{Alessandro Chiolerio}

\address[a]{Unconventional Computing Laboratory, UWE, Bristol, UK}
\address[b]{Department of Engineering and Technology, University of Huddersfield, UK}
\address[c]{Escuela Superior de C\'omputo,  Instituto Polit\'ecnico Nacional, M\'exico}
\address[d]{Democritus University of Thrace, DUTH University Campus, 67100 Xanthi, Greece}
\address[e]{Bioinspired Soft Robotics, Istituto Italiano di Tecnologia, Via Morego 30, 16163 Genova, Italy}

\begin{abstract}
 The colloid cellular automata do not imitate the physical structure of colloids but are governed by logical functions derived from the colloids. We analyse the space-time complexity of Boolean circuits derived from the electrical responses of colloids—specifically ZnO (zinc oxide, an inorganic compound also known as calamine or zinc white, which naturally occurs as the mineral zincite), proteinoids (microspheres and crystals of thermal abiotic proteins), and combinations thereof to electrical stimulation. To extract Boolean circuits from colloids, we send all possible configurations of two-, four-, and eight-bit binary strings, encoded as electrical potential values, to the colloids, record their responses, and thereby infer the Boolean functions they implement. We map the discovered functions onto the cell-state transition rules of cellular automata (arrays of binary state machines that update their states synchronously according to the same rule) --- the colloid cellular automata.  We then analyse the phenomenology of the space-time configurations of the automata and evaluate their complexity using measures such as compressibility, Shannon entropy, Simpson diversity, and expressivity. A hierarchy of phenomenological and measurable space-time complexity is constructed. 
\end{abstract}


\end{frontmatter}

\section{Introduction}

A liquid computer is a device that uses incompressible fluid to process information via mechanical, electrical, optical, or chemical means. The implementation of computation in liquid media has a history spanning over 120 years, from hydraulic algebraic machines developed in the 1900s to fluid maze solvers and droplet logics in the late 2000s. For an overview, please see \cite{adamatzky2019brief}. Advantages of liquid computing include reconfigurability and flexibility, scalability, potential for reduced energy consumption,  bio-compatibility and integration with biological systems, intrinsic parallelism, innovative data storage and retrieval, and novel computation paradigms. While liquid-based computers are still largely experimental and face several technical challenges, they offer intriguing advantages that could revolutionise various fields of computing and technology. 

Recently a new sub-domain of liquid computing emerged -- computing with colloids (mixtures where microscopically dispersed insoluble particles liquids). The rise of colloid computers started from the liquid cybernetic systems, conceptualised as colloidal autonomous soft holonomic processors have demonstrated intriguing features, including autolographic capabilities~\cite{chiolerio2017smart,chiolerio2020liquid}. 
Our previous experiments with ZnO colloids under controlled laboratory conditions demonstrated their potential as electrical analog neurons, successfully implementing synaptic-like learning and Pavlovian reflexes~\cite{kheirabadi2023learning,chiolerio2020liquid}. Additionally, the computational capabilities of $\ce{Fe3O4}$ ferrofluid for digit recognition further exemplify the versatility of liquid-based systems~\cite{crepaldi2023experimental}. 

One of the key developments in colloid computing became mining of Boolean circuits in colloids~\cite{roberts2023logical,fortulan2023reservoir}.
The technique is based on selecting a pair of input sites, applying all possible combinations of inputs, where logical values are represented by electrical characteristics of input signals, to the sites and recording outputs, represented by electrical responses of the substrate, on a set of the selected output sites. The approach belong to the family of reservoir computing~\cite{verstraeten2007experimental,lukovsevivcius2009reservoir,dale2017reservoir,konkoli2018reservoir,dale2019substrate} and   \emph{in materia} computing~\cite{miller2002evolution,miller2014evolution,stepney2019co,miller2018materio,miller2019alchemy} techniques of analysing computational properties of physical and biological substrates. 

In our experimental laboratory studies~\cite{roberts2023logical,fortulan2023reservoir}   we discovered a range of 4-, 6- and 8-ary Boolean functions. In present paper, we evaluate dynamics and complexity of the functions using one-dimensional cellular automata (CA). CA, despite their simple rules and structure, can exhibit complex behaviour. This makes them an excellent tool for evaluating the inherent complexity of $n$-ary Boolean functions by mapping the functions onto the CA rules and observing the resulting dynamics. CA can generate a variety of patterns based on initial states and transition rules. By encoding $n$-Boolean functions into CA rules, we study the patterns that emerge, providing a visual and dynamic representation of the function's complexity. This is particularly useful for understanding how simple functions can lead to complex behaviours and vice versa.

\section{Methods}
\label{methods}

\begin{figure}[!tbp]
    \centering
    \subfigure[]{\includegraphics[scale=0.50]{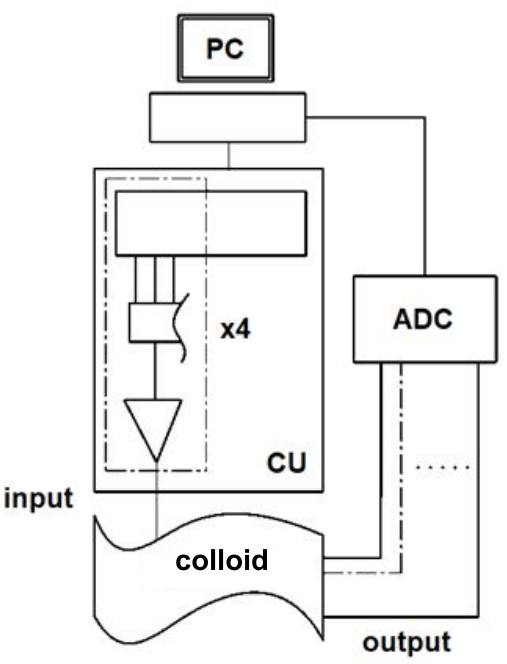}}
    \subfigure[]{\includegraphics[scale=0.08]{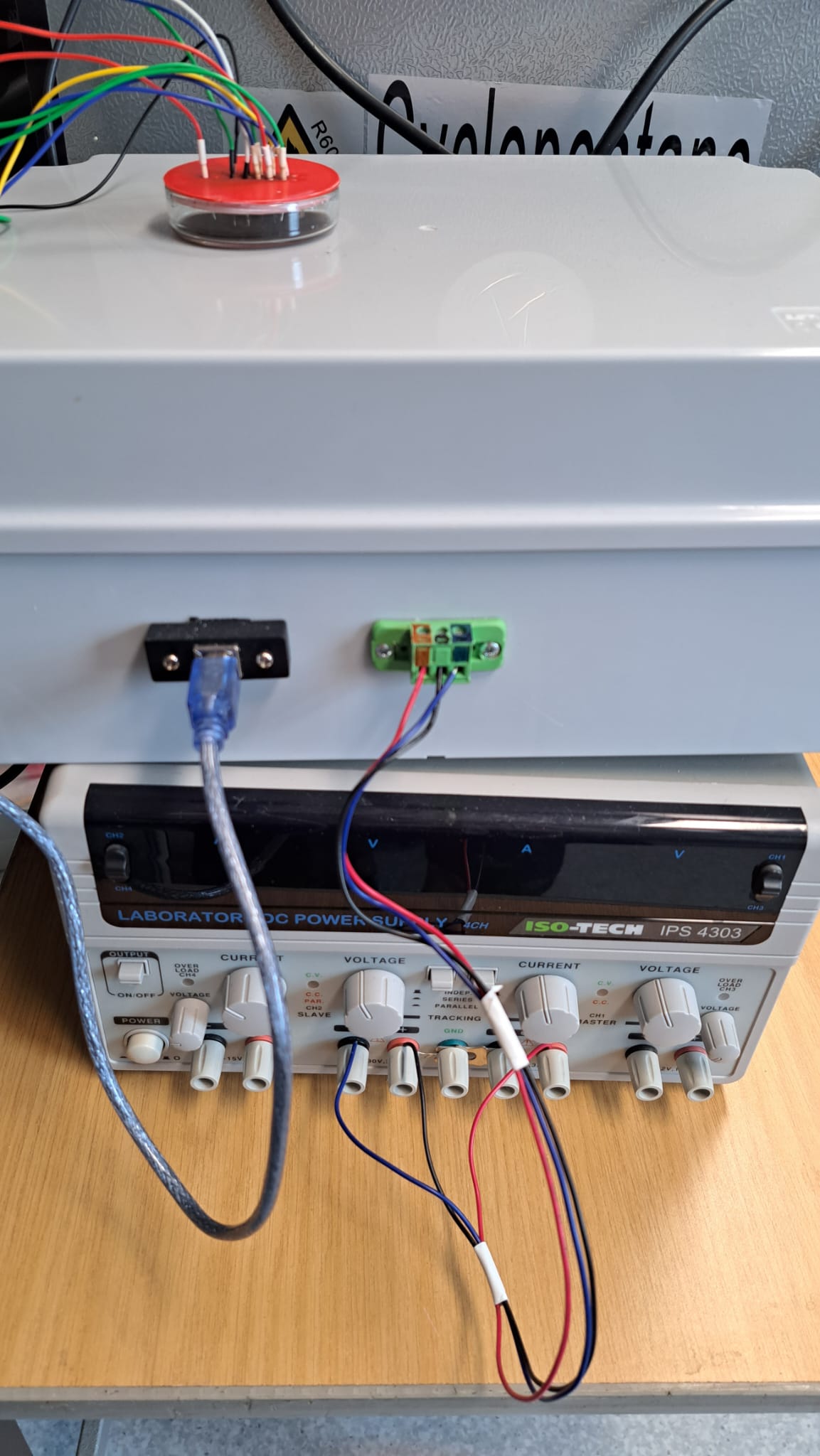}}
    \subfigure[]{\includegraphics[scale=0.40]{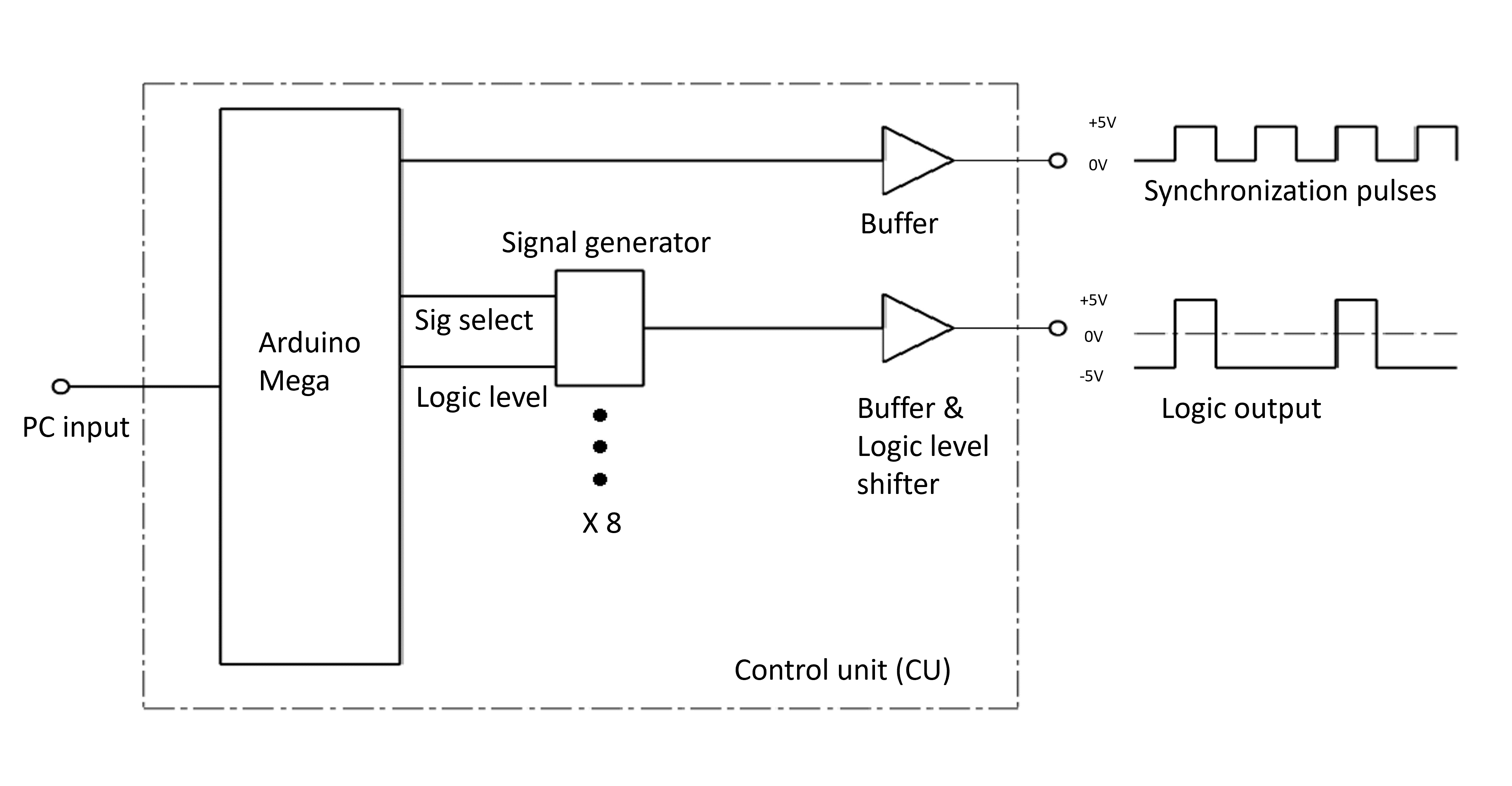}}
    \subfigure[]{\includegraphics[scale=0.20]{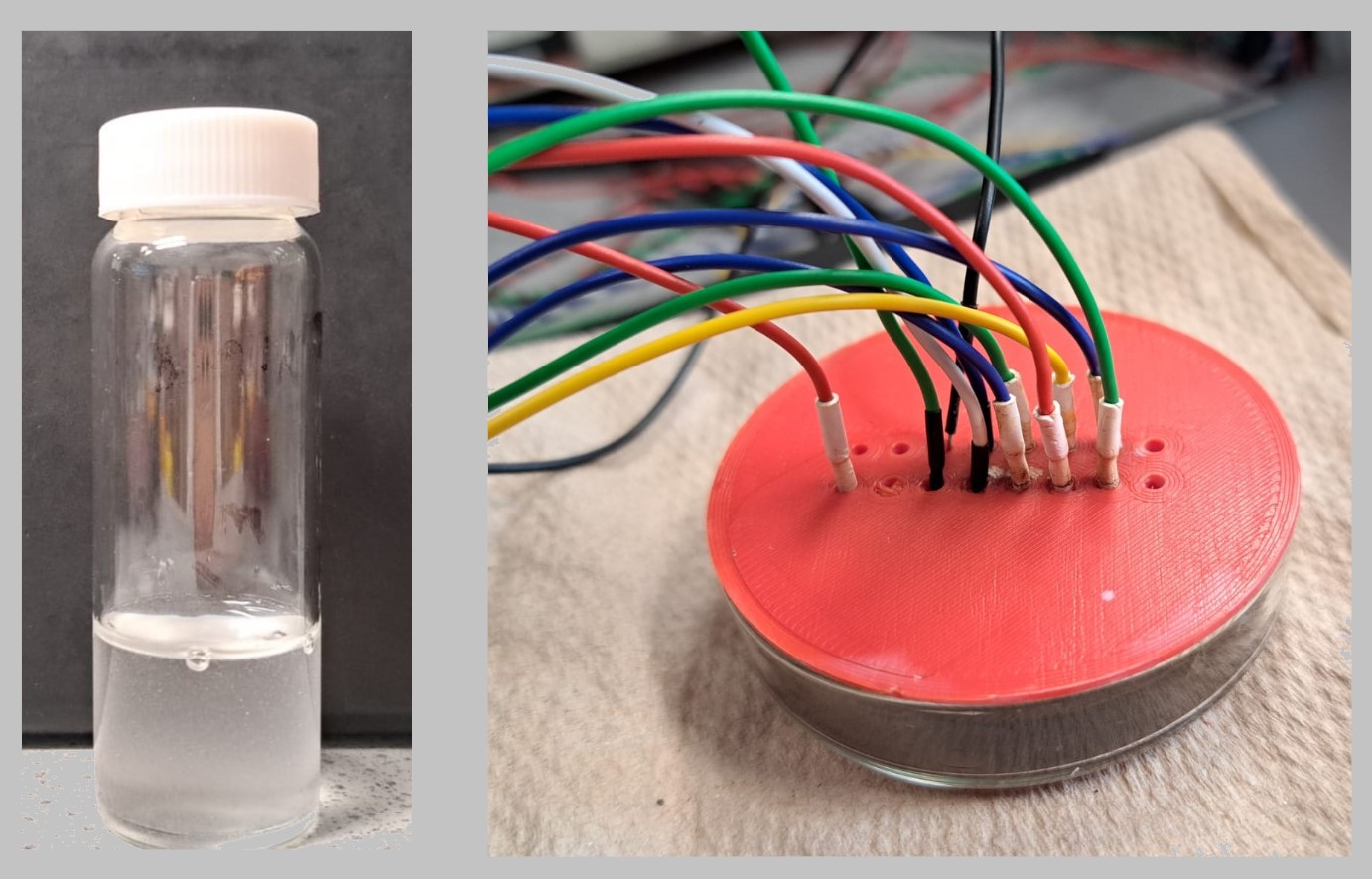}}
    \caption{a) A scheme of the experiments. PC –- laptop for generating sequences; CU -- control unit, the dashed section is a breakdown of a single channel; ADC –- analogue to digital converter~\cite{roberts2023mining}. b) experimental setup. c) A schematic of the inside of the unit control box. d) A close-up photo of the colloid dish and the electrodes interfacing it. From \cite{roberts2023logical}.}
    \label{fig}
\end{figure}

\begin{table}[!tbp]
    \centering
\caption{Most commonly found Boolean functions, $\phi$ is a frequency of the functions' discovery, extracted from ZnO nanoparticle. Boolean functions derived in \cite{roberts2023logical} are shown in (abc), and the functions derived in \cite{fortulan2023reservoir} in (def)
    (ad)~Two-inputs, (be)~Four-inputs. (cf)~Eight inputs.}
\label{tab:functionszno}
{\footnotesize
\subfigure[]{
    \begin{tabular}{l|c}
$f$ & $\phi$ \\ \hline
\\
$f_1=\overline{A} + \overline{B}$ & 73 \\
$f_2=A + B$ & 45\\
$f_3=\overline{A} + B$ & 37\\
$f_4=A + \overline{B}$ & 33\\
$f_5=A\cdot B$ & 8\\
$f_6=B \cdot \overline{A}$ & 6\\
$f_7=(A \cdot \overline{B}) + (B \cdot \overline{A})$ & 4\\
$f_8=(A \cdot B) + (\overline{A} \cdot \overline{B})$ & 3\\
$f_9=A \cdot \overline{B}$ & 3\\
$f_{10}=\overline{A} \cdot \overline{B}$ & 2\\
    \end{tabular}
    }}
{\footnotesize
\subfigure[]{
   \begin{tabular}{l|c}
$f$ & $\phi$ \\ \hline
\\
$f_{11}=(A \cdot \overline{B}) + (B \cdot \overline{A} \cdot \overline{C}) + (B \cdot \overline{C} \cdot \overline{D})$ & 	 7 \\
$f_{12}=(C \cdot D \cdot \overline{B}) + (A \cdot \overline{B} \cdot \overline{D}) + (B \cdot \overline{A} \cdot \overline{D}) + (D \cdot \overline{A} \cdot \overline{C})$ & 	 6 \\
$f_{13}=(A \cdot \overline{B} \cdot \overline{D}) + (B \cdot \overline{A} \cdot \overline{C} \cdot \overline{D})$ & 	 6 \\
$f_{14}=(\overline{A} \cdot \overline{D}) + (A \cdot B \cdot C \cdot D) + (B \cdot \overline{A} \cdot \overline{C}) + (C \cdot \overline{A} \cdot \overline{B})$ & 	 5 \\
$f_{15}=(A \cdot \overline{B} \cdot \overline{D}) + (B \cdot \overline{A} \cdot \overline{C}) + (B \cdot \overline{C} \cdot \overline{D})$ & 	 5 \\
$f_{16}=A \cdot D \cdot \overline{B} \cdot \overline{C}$ & 	 5 \\
$f_{17}=A \cdot \overline{B} \cdot \overline{C} \cdot \overline{D}$ & 	 5 \\
$f_{18}=(B \cdot C \cdot D) + (B \cdot C \cdot \overline{A}) + (C \cdot D \cdot \overline{A}) + (A \cdot \overline{B} \cdot \overline{C} \cdot \overline{D})$ & 	 5 \\ 
$f_{19}=(A \cdot D \cdot \overline{B}) + (B \cdot D \cdot \overline{A}) + (A \cdot \overline{B} \cdot \overline{C}) + (B \cdot \overline{A} \cdot \overline{C}) + (D \cdot \overline{A} \cdot \overline{C})$ & 	 5 \\
$f_{20}=(D \cdot \overline{A}) + (D \cdot \overline{B}) + (B \cdot \overline{A} \cdot \overline{C})$ & 	 5 \\
    \end{tabular}
}}
{\footnotesize
\subfigure[]{
\begin{tabular}{p{15cm}}
$f_{21}=(A \cdot B \cdot F \cdot \overline{C} \cdot \overline{E})+(A \cdot D \cdot F \cdot \overline{C} \cdot \overline{E})+(A \cdot G \cdot H \cdot \overline{B} \cdot \overline{C})+(B \cdot D \cdot E \cdot \overline{A} \cdot \overline{F})+(B \cdot E \cdot H \cdot \overline{A} \cdot \overline{C})+(C \cdot D \cdot E \cdot \overline{B} \cdot \overline{F})+(D \cdot E \cdot H \cdot \overline{B} \cdot \overline{G})+(D \cdot F \cdot H \cdot \overline{A} \cdot \overline{B})+(B \cdot C \cdot D \cdot F \cdot G \cdot \overline{H})+(B \cdot C \cdot D \cdot G \cdot H \cdot \overline{F})+(B \cdot E \cdot \overline{A} \cdot \overline{G} \cdot \overline{H})+(C \cdot F \cdot \overline{B} \cdot \overline{G} \cdot \overline{H})+(E \cdot H \cdot \overline{A} \cdot \overline{C} \cdot \overline{F})+(F \cdot H \cdot \overline{B} \cdot \overline{D} \cdot \overline{E})+(A \cdot C \cdot E \cdot G \cdot \overline{B} \cdot \overline{D})+(A \cdot E \cdot F \cdot G \cdot \overline{C} \cdot \overline{D})+(B \cdot D \cdot E \cdot G \cdot \overline{C} \cdot \overline{F})+(B \cdot E \cdot F \cdot G \cdot \overline{A} \cdot \overline{D})+(C \cdot F \cdot G \cdot H \cdot \overline{D} \cdot \overline{E})+(A \cdot C \cdot D \cdot E \cdot F \cdot H \cdot \overline{G})+(B \cdot \overline{C} \cdot \overline{E} \cdot \overline{G} \cdot \overline{H})+(C \cdot \overline{B} \cdot \overline{D} \cdot \overline{F} \cdot \overline{G})+(D \cdot \overline{C} \cdot \overline{E} \cdot \overline{F} \cdot \overline{H})+(E \cdot \overline{A} \cdot \overline{B} \cdot \overline{D} \cdot \overline{H})+(E \cdot \overline{B} \cdot \overline{D} \cdot \overline{G} \cdot \overline{H})+(B \cdot C \cdot D \cdot \overline{A} \cdot \overline{E} \cdot \overline{G})+(B \cdot D \cdot F \cdot \overline{C} \cdot \overline{G} \cdot \overline{H})+(B \cdot D \cdot G \cdot \overline{A} \cdot \overline{C} \cdot \overline{E})+(B \cdot E \cdot H \cdot \overline{C} \cdot \overline{F} \cdot \overline{G})+(C \cdot D \cdot G \cdot \overline{B} \cdot \overline{E} \cdot \overline{H})+(C \cdot E \cdot F \cdot \overline{D} \cdot \overline{G} \cdot \overline{H})+(C \cdot G \cdot H \cdot \overline{A} \cdot \overline{E} \cdot \overline{F})+(D \cdot E \cdot F \cdot \overline{A} \cdot \overline{B} \cdot \overline{C})+(B \cdot D \cdot \overline{E} \cdot \overline{F} \cdot \overline{G} \cdot \overline{H})+(B \cdot F \cdot \overline{A} \cdot \overline{D} \cdot \overline{E} \cdot \overline{G})+(C \cdot D \cdot \overline{A} \cdot \overline{B} \cdot \overline{E} \cdot \overline{H})+(C \cdot H \cdot \overline{D} \cdot \overline{E} \cdot \overline{F} \cdot \overline{G})+(A \cdot C \cdot E \cdot G \cdot \overline{D} \cdot \overline{F} \cdot \overline{H})+(A \cdot \overline{B} \cdot \overline{C} \cdot \overline{F} \cdot \overline{G} \cdot \overline{H})+(D \cdot \overline{A} \cdot \overline{B} \cdot \overline{C} \cdot \overline{E} \cdot \overline{G})+(G \cdot \overline{A} \cdot \overline{C} \cdot \overline{D} \cdot \overline{E} \cdot \overline{H})$
\end{tabular}
}}
{\footnotesize
\subfigure[]{
		\begin{tabular}{l|c}
			Function & $\phi$   \\ \hline
			$f_{22}=A+ B$  & 35      \\
			$f_{23}=A\cdot B$ & 3       \\
		\end{tabular}
  }}
{\footnotesize
\subfigure[]{
		\begin{tabular}{l|c}
			$f$ & $\phi$  \\ \hline
			$f_{24}=A + B + C + D$  & 16     \\
			$f_{25}=(A \cdot B \cdot D \cdot \overline{ C}) + (C \cdot D \cdot \overline{  A} \cdot \overline{ B})$    & 6      \\ 
			$f_{26}=(C \cdot D) + (A \cdot B \cdot D)$                                                                         & 4      \\ 
			$f_{27}=A + B + D$                                                     & 2   
		\end{tabular}}}
{\footnotesize
\subfigure[]{
			\begin{tabular}{l|c}
				$f$ & $\phi$  \\ \hline
				$f_{28}=\overline{ A} + \overline{ B} + \overline{ C} + \overline{ D} + \overline{ E} + \overline{ F} + \overline{ G}$                                     & 4 \\
				$f_{29}=A \cdot C \cdot D \cdot E \cdot F \cdot G \cdot H \cdot \overline{ B}$              & 2 \\ 
				$f_{30}=A + C + D + E + F + H + (B \cdot \overline{ G}) + (G \cdot \overline{ B})$                                                                               & 1 \\ 
				$f_{31}=C + D + E + F + H + (A \cdot B) + (A \cdot \overline{ G}) + (B \cdot \overline{ G}) + (G \cdot \overline{ A} \cdot \overline{ B})$   & 1\\
			\end{tabular}}}
\end{table}

\begin{table}[!bp]
		\centering
		\caption{The four most common extracted sum-of-products Boolean expressions with varying thresholds for the dispersed proteinoids (abc) and mixture of ZnO and proteinoids (def), (ad)~2-bit input string, (be)~4-bit input string, (cf)~8-bit input string.}
		\label{tab:sop_2bit_prot}
  {\footnotesize
\subfigure[]{
		\begin{tabular}{l|c}
			$f$  & $\phi$   \\ \hline
			$0\quad(\mathrm{False})$                     & 19     \\
			$f_{22}=A+ B$                                                        & 16       \\ 
			$f_{23}=A\cdot B$                                                  & 3       \\ 
		\end{tabular}}}
{\footnotesize
\subfigure[]{
		\begin{tabular}{l|c}
			$f$ & $\phi$   \\ \hline
			$f_{24}=A + B + C + D$  & 23     \\
			$f_{32}=A \cdot B \cdot C \cdot D$ & 4 \\ 
			$f_{33}=A + B + (C \cdot D)$ & 3 \\ 
			$f_{34}=(A \cdot B) + (B \cdot D) + (C \cdot D) + (A \cdot \overline{C} \cdot \overline{ D}) $ & 3\\
		\end{tabular}}}
  {\footnotesize
\subfigure[]{
			\begin{tabular}{l|c}
				$f$ & Count \\ \hline
				$f_{35}=\overline{ A} + \overline{ B} + \overline{ C} + \overline{ D} + \overline{ E} + \overline{ F} + \overline{ G} + \overline{ H}$ & 19    \\
				$f_{36}=(A \cdot \overline{ E}) + (B \cdot \overline{ H}) + (C \cdot \overline{ G}) + (D \cdot \overline{ F}) + (E \cdot \overline{ D}) + (F \cdot \overline{ C}) + (G \cdot \overline{ B}) + (H \cdot \overline{ A})$ & 4     \\
				$f_{37}=A \cdot B \cdot C \cdot D \cdot E \cdot F \cdot H \cdot \overline{ G }\cdot $ & 2     \\
				\begin{tabular}[c]{@{}l@{}}$(C \cdot \overline{ B}) + (C \cdot \overline{ D}) + (D \cdot \overline{ E}) + (E \cdot \overline{ G}) + (F \cdot \overline{ H}) + (G \cdot \overline{ F}) \lor$ \\ $f_{38}=(H \cdot \overline{ E}) + (A \cdot B \cdot \overline{ C}) + (A \cdot H \cdot \overline{ B}) + (B \cdot H \cdot \overline{ C}$\end{tabular} & 1    
		\end{tabular}}}
  {\footnotesize
\subfigure[]{
		\begin{tabular}{l|c}
			$f$ & $\phi$   \\ \hline
			$f_{23}=A\cdot B$                & 21     \\
			$f_{22}=A+ B$                   & 14     \\ 
			$B$                             & 2      \\
			$A$                             & 1       \\ 
		\end{tabular}}}
  {\footnotesize
\subfigure[]{
		\begin{tabular}{l|c}
			$f$ & $\phi$ \\ \hline
			$f_{39}=\overline{ A} + \overline{ B} + \overline{ C}$& 15   \\
			$f_{40}=(A \cdot \overline{ B}) + (B \cdot \overline{ C}) + (D \cdot \overline{ A})$ & 7    \\
		$f_{41}=A \cdot B \cdot D \cdot \overline{ C}$&5    \\
			$f_{42}=(A \cdot B \cdot \overline{ C}) + (A \cdot D \cdot \overline{B}) + (B \cdot D \cdot \overline{ C})$  & 4    \\ 
		\end{tabular}}}
{\footnotesize
\subfigure[]{
			\begin{tabular}{l|c}
				$f$ & $\phi$ \\ \hline
				$f_{43}= A +  B + C +  D +  E +  F +  G +  H$& 6\\
				$f_{44}=A \cdot B \cdot C \cdot D \cdot E \cdot F \cdot G \cdot \overline{ H}$& 3\\
				$f_{45}=(A \cdot \overline{ D}) + (B \cdot \overline{ G}) + (C \cdot \overline{ F}) + (D \cdot \overline{ E}) + (E \cdot \overline{ C}) + (F \cdot \overline{ B}) + (G \cdot \overline{ A}) + (G \cdot \overline{ H})$& 2\\
				$f_{46}=(A + C + D + E + (B \cdot F) + (B \cdot G) + (B \cdot H) + (F \cdot G) + (F \cdot H) + (G \cdot H)$& 1\\ 
		\end{tabular}}}
	\end{table}

 Experimental techniques on mining Boolean functions are described in full details in~\cite{roberts2023logical,fortulan2023reservoir}. Here we briefly outline an overall approach, based on the example of \cite{roberts2023logical}.
Colloids of ZnO and proteinoids have been prepared as detailed in~\cite{roberts2023logical,fortulan2023reservoir}. 
The hardware was built around an Arduino Mega 2560 (Elegoo, China) and a series of AD9833 programmable signal generators (Analog, USA). This setup can send sequences of 2, 4, and 8-bit strings to the colloid sample, with the strings encoded as step voltage inputs: -5 V representing a logical `0' and 5 V representing a logical `1'.  In Fig. \ref{fig}(a), a PC programs a Control Unit (CU) and receives readings from an analog-to-digital converter (ADC). The CU, shown as a grey box connected to a standard laboratory power supply in Fig. \ref{fig}(b), contains the Arduino Mega and multiple amplifiers. To generate the 2, 4, and 8-bit strings without redesigning or rewiring the CU, multiple programmable signal generators were incorporated. This is abstracted in Fig. \ref{fig}(c), where only one generator and its output are depicted for simplicity. Activation of these generators is controlled by the Arduino Mega, which is programmed through the PC and also depicted within the CU entity in Fig. \ref{fig}(c).
To search for 2-, 4-, and 8-input Boolean circuits, we used respective electrodes. These were 10 µm platinum rods inserted 5 mm apart into the colloid container. Data acquisition (DAQ) probes, separated by 5 mm, fed 2 differential outputs to a Pico 24 ADC. Its 3\textsuperscript{rd} channel received a pulse on each input state change. Refer to Fig.~\ref{fig} for the apparatus schematic. The strings counted from binary 00 to 11, 0000 to 1111, or 00000000 to 11111111, changing state every 15 seconds. All possible electrode states were tested. For two bits, states sequentially altered every 15 seconds between 00, 01, 10, and 11. Similarly, all states of the four- and eight-bit strings were sequentially applied.  Samples from 2 channels were taken at 1 Hz throughout the experiment. Peaks for each channel were located for 10 thresholds, from 100 mV to 600 mV in 50 mV steps, for each input state from 0000 to 1111. 
Most commonly found Boolean functions extracted from ZnO nanoparticle are listed in Tab.~\ref{tab:functionszno}.  
Boolean functions derived in \cite{roberts2023logical} are presented in  Tab.~\ref{tab:functionszno}(abc), and the functions derived in \cite{fortulan2023reservoir} in Tab.~\ref{tab:functionszno}(def). Most frequent Boolean functions discovered in proteinoid colloids are shown in 
Tab.~\ref{tab:sop_2bit_prot}(abc) and mixture of ZnO and proteinoids in Tab.~\ref{tab:sop_2bit_prot}(def).

We evaluate complexity of the functions discovered  via complexity of the space-time configurations of one-dimensional cellular automata (CA)
governed by these functions. We call these CA `colloid cellular automata' because their space-time evolution is governed by Boolean functions implemented by colloids and their mixtures in laboratory experiments.  We consider an array $Z$ of finite state automata, called cells, where every cell takes states `0' or `1' and updates its state depending on the states of its two, four or eight immediate neighbours. All cells update their states by the same rule and in discrete time. For example, a cell with index $i$, $x_i \in Z$, updates its state at time $t$ as a function of states of its two neighbours $x^{t+1}=f(x_{i-1}^t, x_{i-2}^t)$ (representing variables $A$ and $B$ in  Tab.~\ref{tab:functionszno}ad  and Tab.~\ref{tab:sop_2bit_prot}ad),
four neighbours: $x^{t+1}=f(x_{i-2}^t, x_{i-1}^t, x_{i+1}^t, x_{i+2}^t)$ (representing variables $A \ldots C$ in  Tab.~\ref{tab:functionszno}be and Tab.~\ref{tab:sop_2bit_prot}be), or eight neighbours
$x^{t+1}=f(x_{i-4}^t, x_{i-3}^t, x_{i-2}^t, x_{i-1}^t, x_{i+1}^t, x_{i+2}^t, x_{i+3}^t, x_{i+4}^t)$ (representing variables $A \ldots H$ in  Tab.~\ref{tab:functionszno}cf  and Tab.~\ref{tab:sop_2bit_prot}cf). For example the function $f_{25}=(A \cdot B \cdot D \cdot \overline{ C}) + (C \cdot D \cdot \overline{  A} \cdot \overline{ B})$ is represented in a cell-state transition rule as 
$x^{t+1}=(x_{i-2}^t \cdot x_{i-1}^t \cdot x_{i+2}^t \cdot \overline{x_{i+1}^t}) + (x_{i+1}^t \cdot x_{i+2}^t \cdot \overline{x_{i-2}^t} \cdot \overline{x_{i-1}^t})$. We evolved automata of 500 cells in 500 iterations of simultaneous cell-state transition. 

To evaluate complexity of the cellular automata we used Shannon entropy~\cite{shannon1948mathematical,lin1991divergence,eskov2017shannon}, Simpson's diversity (commonly used in ecological studies to evaluate biodiversity of populations~\cite{somerfield2008simpson,nagendra2002opposite,kim2017deciphering}), Lempel-Ziv complexity~\cite{ziv1977universal}, space filling and expressiveness~\cite{adamatzky2012phenomenology,redeker2013expressiveness}.
Let matrix $L$ represent time-space configuration of a 1D CA governed by state-transition rules derived from colloids.
Let  $W=\{ 0,1 \}^9$ be a set of all possible configurations of a 9-node neighbourhood $B_x$ including the central node $x$. 
 Let $B$ be a configuration of matrix $L$, we calculate a number of non-quiescent  neighbourhood 
configurations as $\eta = \sum_{ x \in L} \epsilon(x)$, 
where $\epsilon(x)=0$ if for every resting $x$ all 
its neighbours are in the state `0', and $\epsilon(x)=1$ otherwise. 
The Shannon entropy $H$ is calculated as 
$H =- \sum_{w \in W} (\nu(w)/\eta \cdot ln (\nu(w)/\eta))$, 
where $\nu(w)$ is a number of times the neighbourhood 
configuration $w$ is found in configuration $B$. 
Simpson's  diversity $S$ is calculated as $S=\sum_{w \in W} (\nu(w)/\eta)^2$. 
Simpson diversity linearly correlates with Shannon entropy for $H<3$; relationships becomes logarithmic for higher values of $H$ as we previously demonstrated in \cite{adamatzky2018generative}.
The assessment of Lempel-Ziv complexity (compressibility), denoted as $LZ$, is based on the size of space-time configurations saved as PNG files representing configurations. This approach suffices because the 'deflation' algorithm utilised in PNG lossless compression, as outlined in \cite{roelofs1999png,howard1993design,deutsch1996zlib}, is a derivative of the classical Lempel--Ziv 1977 algorithm, as described in \cite{ziv1977universal}.
Space filling $D$ is a ratio of non-zero entries in $B$ to the total number of cells/nodes. This is used to estimate expressiveness. 
Expressiveness $E$ is calculated as the Shannon entropy $H$ divided by space-filling ratio $D$, the expressiveness reflects the `economy of diversity'.

\section{Results}

\begin{figure}[!tbp]
    \centering
\subfigure[$f_2$]{
    \includegraphics[width=0.3\textwidth]{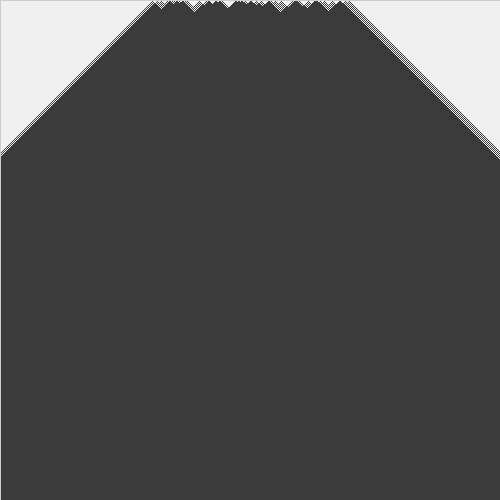}}
    \subfigure[$f_1, f_{10}$]{
    \includegraphics[width=0.3\textwidth]{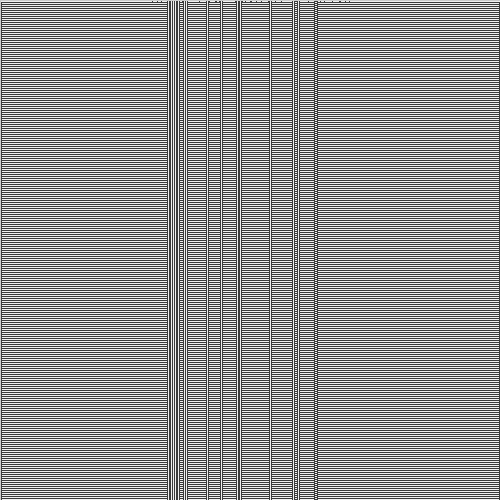}}
\subfigure[$f_3,f_6, f_{38}$]{
    \includegraphics[width=0.3\textwidth]{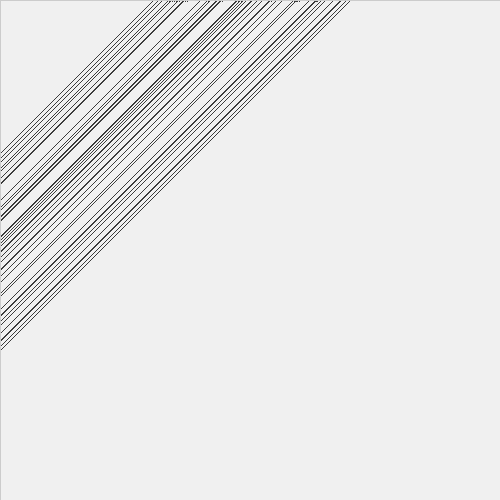}}
\subfigure[$f_4, f_9, f_{11}, f_{13}, f_{15}, f_{17}$, $f_{18}$]{
    \includegraphics[width=0.3\textwidth]{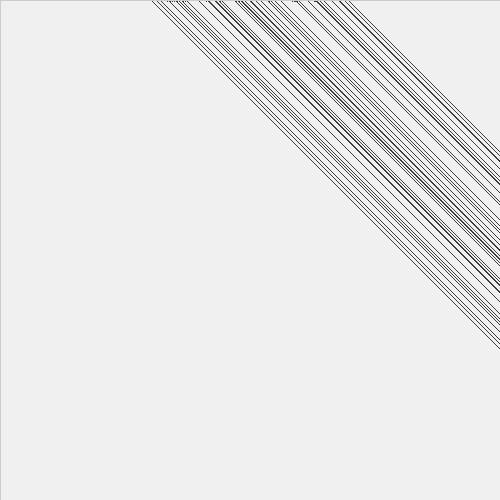}}
\subfigure[$f_7$]{
    \includegraphics[width=0.3\textwidth]{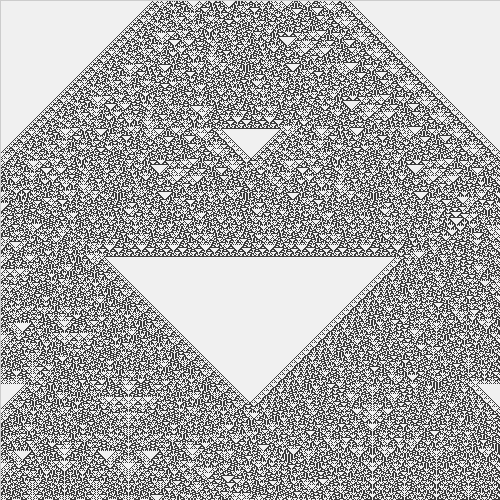}}
\subfigure[$f_8$]{
    \includegraphics[width=0.3\textwidth]{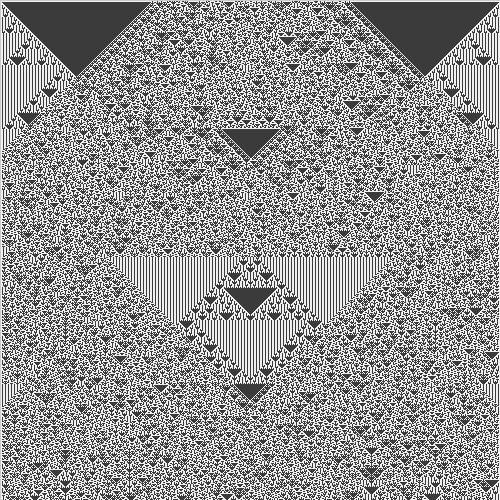}}
\subfigure[$f_{39}, f_{28}, f_{35}$]{
    \includegraphics[width=0.3\textwidth]{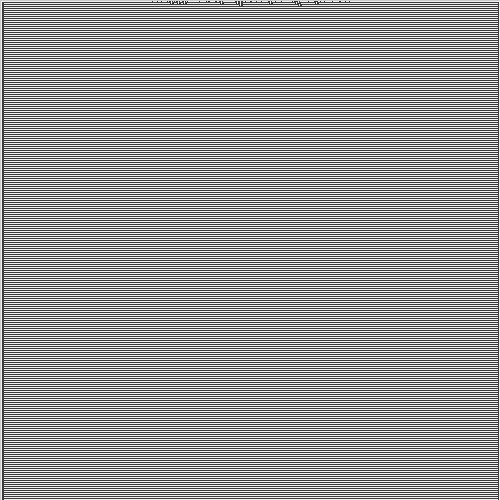}}
\subfigure[$f_{40}$]{
    \includegraphics[width=0.3\textwidth]{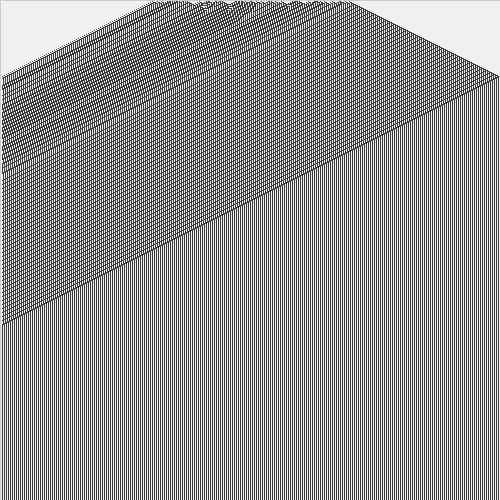}}
    \caption{Functions with two-arguments and those functions with four or eight arguments which produce alike patterns. Space (1D CA array) states are horizontal, and time (progressing from top to bottom) is vertical:\\
    $x^0_1 x^0_2 \ldots x^0_{500}$\\
    $x^1_1 x^1_2 \ldots x^1_{500}$\\
    $\ldots$\\
    $x^{500}_1 x^{500}_2 \ldots x^{500}_{500}$\\
    }
    \label{fig:binaryneighbourhod}
\end{figure}

\begin{figure}[!h]
    \centering
    \subfigure[$f_{12}$]{
    \includegraphics[width=0.3\textwidth]{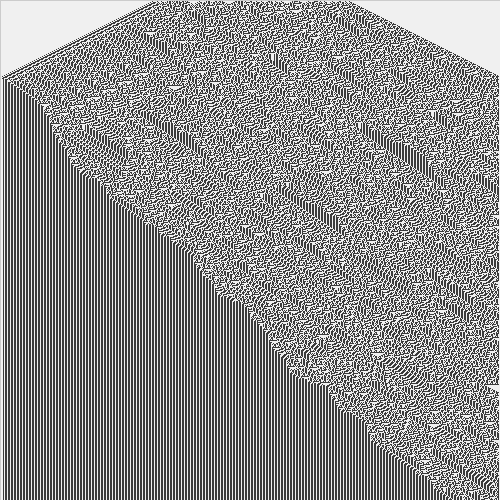}}
    \subfigure[$f_{14}$]{
    \includegraphics[width=0.3\textwidth]{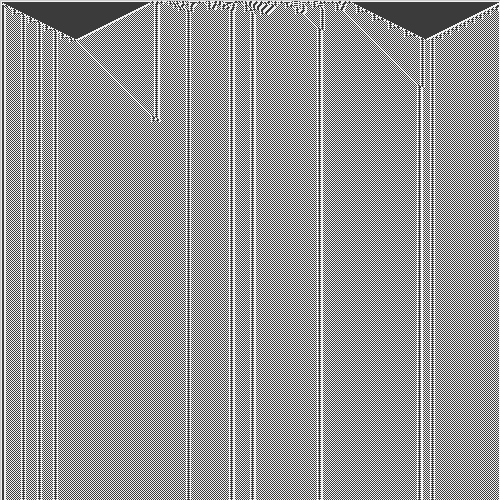}}\\
    \subfigure[$f_{19}$]{
    \includegraphics[width=0.3\textwidth]{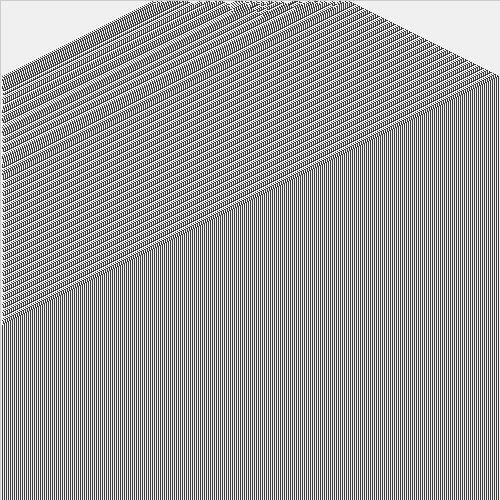}}
    \subfigure[$f_{20}$]{
    \includegraphics[width=0.3\textwidth]{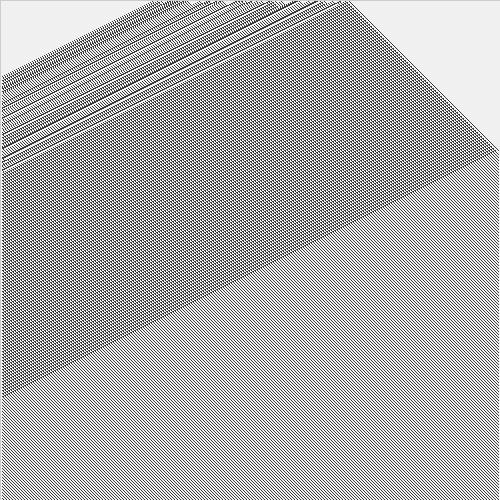}}
    \caption{Functions with four-arguments and those functions with eight arguments which produce alike patterns. Space (1D CA array) states are horizontal, and time (progressing from top to bottom) is vertical.}
    \label{fig:4neighbourhod}
\end{figure}

\begin{figure}[!h]
    \centering
    \subfigure[$f_{21}$]{
    \includegraphics[width=0.3\textwidth]{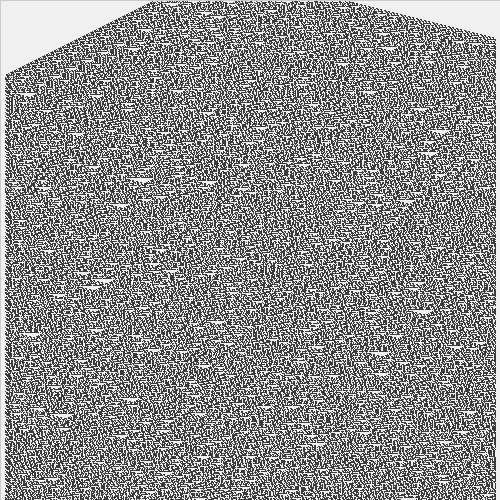}}
    \subfigure[$f_{36}, f_{45}$]{
    \includegraphics[width=0.3\textwidth]{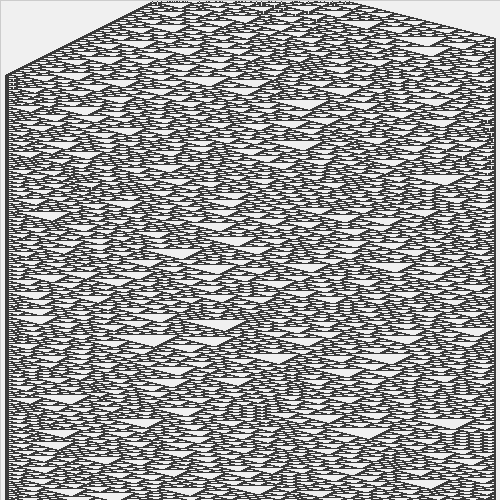}}
    \subfigure[$f_{37}$]{
    \includegraphics[width=0.3\textwidth]{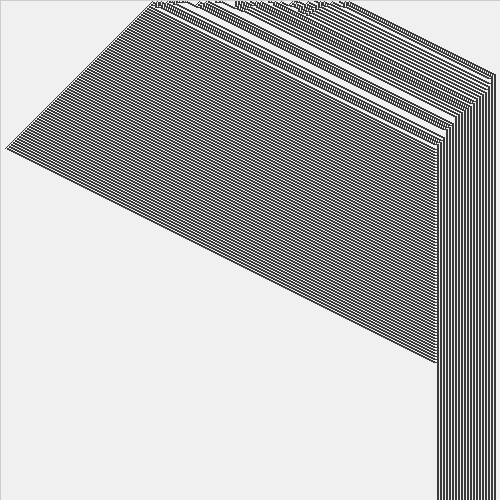}}
\caption{Functions with eight-arguments. Space (1D CA array) states are horizontal, and time (progressing from top to bottom) is vertical.}
    \label{fig:8neighbourhod}
\end{figure}

CA presented by majority of functions from Tabs.~\ref{tab:functionszno} and \ref{tab:sop_2bit_prot} evolve to all-0 or all-1 state, an example of evolution to all-0 states is shown in Fig.~\ref{fig:binaryneighbourhod}a. These are `trivial' functions. Let us consider the positions of the functions within Wolfram's classification of CA behaviour~\cite{wolfram1983statistical}. Most functions discovered belong to Class I, which is characterised by automata exhibiting simple dynamics and evolving to a stable state where all cells are in the same state. 
Functions $f_1$, $f_{10}$ (Fig.~\ref{fig:binaryneighbourhod}b),
$f_{39}$, $f_{28}$, $f_{35}$ (Fig.~\ref{fig:binaryneighbourhod}g), $f_{40}$ (Fig.~\ref{fig:binaryneighbourhod}h), $f_{12}$ (Fig.~\ref{fig:4neighbourhod}a), $f_{14}$ (Fig.~\ref{fig:4neighbourhod}b), $f_{19}$ (Fig.~\ref{fig:4neighbourhod}c), $f_{20}$ (Fig.~\ref{fig:4neighbourhod}d), $f_{37}$ (Fig.~\ref{fig:8neighbourhod}c), 
$f_3$, $f_6$, $f_{38}$  (Fig.~\ref{fig:binaryneighbourhod}c), $f_4$, $f_9$, $f_{13}$, $f_{15}$, $f_{17}$, $f_{18}$ (Fig.~\ref{fig:binaryneighbourhod}d)
and the function $f_{37}$ (Fig.~\ref{fig:8neighbourhod}c)
belong to the class II: the automata fall into global cells do not update their state or update them cyclically from `0' to `1'. 
Space-time dynamics of class III CA is  by quasi-random behaviour and difficult predictability of the successions of the global states. The following functions can be related to the class III CA: $f_7$ (Fig.~\ref{fig:binaryneighbourhod}e), $f_8$ (Fig.~\ref{fig:binaryneighbourhod}f), $f_{21}$ (Fig.~\ref{fig:8neighbourhod}a), $f_{36}$ and $f_{45}$ (Fig.~\ref{fig:8neighbourhod}b). No functions from those discovered in laboratory experiments seem to belong to class IV, where the space-time dynamics of automata show gliders (compact patterns translating in space) with non-trivial interactions between the gliders. CA governed by functions presented in Fig.~\ref{fig:binaryneighbourhod}cd demonstrate travelling compact patterns however these patterns emerge due to asymmetry of the functions.

\begin{table}[!tbp]
    \centering
 \caption{Complexity of space-time patterns generated by CA derived from non-trivial Boolean functions mined in ZnO and proteinoids' colloids:  $LZ$ is an LZ complexity measured via size of ZIP file of the space-time configurations, $LZ/n$ is the complexity normalised by the input string size, $H$ is Shannon entropy, $S$ is Simpson diversity, $D$ is a space filling, $E$ is an expressiveness. (a)~Two-arguments functions, (b)~Four-arguments functions, (c)~Eight-arguments functions. }
    \label{tab:complexity}
{\footnotesize
    \subfigure[]{
    \begin{tabular}{c|c|c|c|c|c|c}
$f$	&	$LZ$	&	$LZ/n$	&	$H$	&	$S$	&	$D$	&	$E$ \\ \hline
$f_2$	&	4	&	2	&	0.1	&	0.02	&	0.98	&	0.1 \\
$f_{11}$	&	9	&	4.5	&	0.07	&	0.02	&	0.98	&	0.1 \\
$f_3$	&	14	&	7	&	0.05	&	0.03	&	0.1	&	0.5 \\
$f_4$	&	14	&	7	&	0.05	&	0.03	&	0.1	&	0.5 \\
$f_1$	&	16	&	8	&	0.5	&	0.2	&	0.9	&	0.6 \\
$f_7$	&	57	&	28.5	&	1.9	&	0.8	&	0.4	&	4.8 \\
$f_8$	&	61	&	30.5	&	1.9	&	0.8	&	0.5	&	3.8 \\
    \end{tabular} }  }
{\footnotesize
    \subfigure[]{
    \begin{tabular}{c|c|c|c|c|c|c}
$f$	&	$LZ$	&	$LZ/n$	&	$H$	&	$S$	&	$D$	&	$E$\\ \hline
$f_{12}$	&	11	&	2.75	&	1.4	&	0.7	&	0.6	&	2.3 \\
$f_{14}$	&	12	&	3	&	1.1	&	0.6	&	0.5	&	2.2 \\
$f_{19}$	&	22	&	5.5	&	0.7	&	0.5	&	0.5	&	1.4 \\
$f_{20}$	&	42	&	10.5	&	1.1	&	0.7	&	0.32	&	3.4 \\
$f_{40}$	&	9	&	4.5	&	0.7	&	0.5	&	0.5	&	1.4 \\
\end{tabular} }  }
{\footnotesize
    \subfigure[]{
    \begin{tabular}{c|c|c|c|c|c|c}
$f$	&	$LZ$	&	$LZ/n$	&	$H$	&	$S$	&	$D$	&	$E$\\ \hline
$f_{37}$	&	11	&	1.375	&	1.2	&	0.7	&	0.1	&	12.0	\\
$f_{36}$	&	36	&	4.5	&	1.1	&	0.5	&	0.5	&	2.2	\\
$f_{21}$	&	65	&	8.125	&	1.9	&	0.8	&	0.5	&	3.8	\\
\end{tabular} }  }
\end{table}

\begin{figure}[!tbp]
    \centering
\subfigure[]{
    \includegraphics[width=0.47\textwidth]{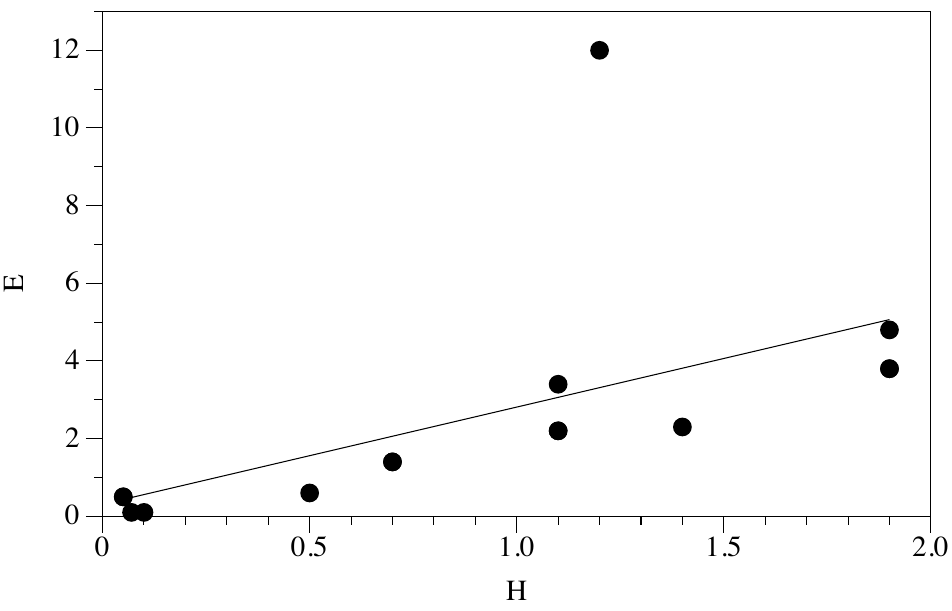}
    }
\subfigure[]{
    \includegraphics[width=0.47\textwidth]{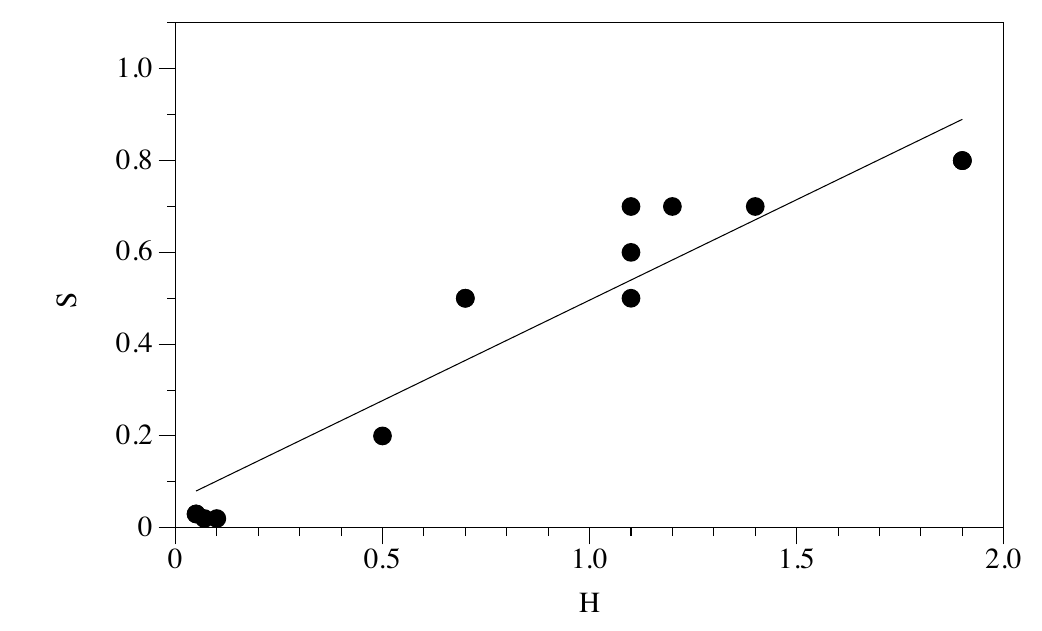}
    }
\subfigure[]{
    \includegraphics[width=0.47\textwidth]{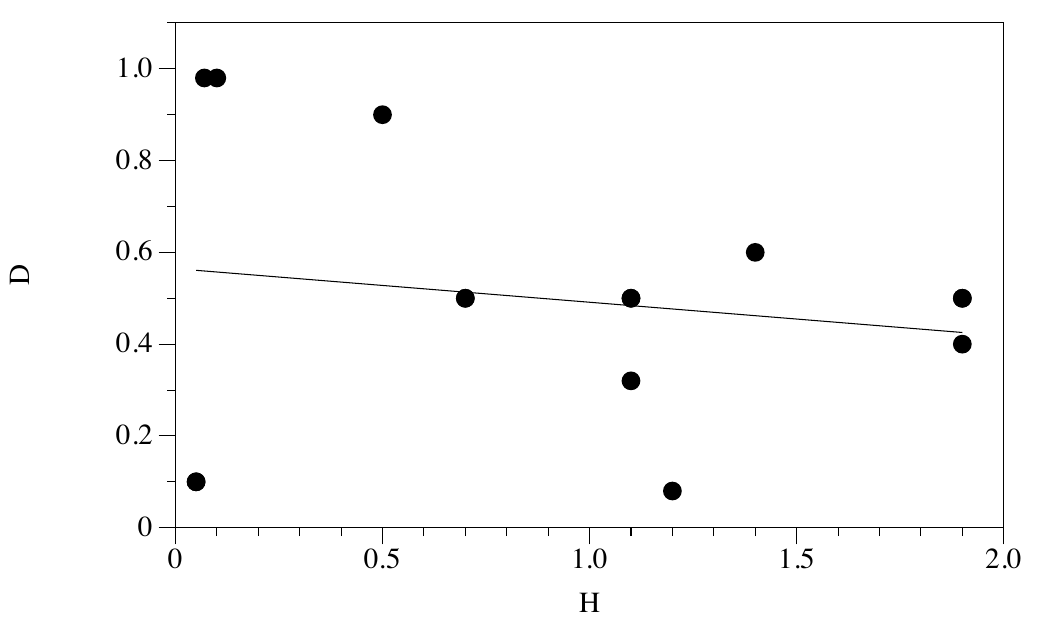}
    }
    \caption{(a)~Shannon entropy $H$ vs expressiveness $E$, linear approximation $E=0.30874 + 2.5032*H$. 
    (b)~Shannon entropy $H$ vs Simpson index $S$, linear approximation $S=0.058092 + 0.43781*H$.
    (c)~Shannon entropy $H$ vs space filling $D$, linear approximation $D=0.56455 + (-0.073223)*H$.}
    \label{fig:scatter}
\end{figure}

Complexity measures of the functions discussed are shown in Tab.~\ref{tab:complexity}. Complexity measures --- Shannon entropy, Simpson index and expressiveness -- are  consistent with each other as seen in scatter plots for Shannon entropy $H$ vs. expressiveness $E$ (Fig.~\ref{fig:scatter}a), 
Shannon entropy $H$ vs.  Simpson index $S$ (Fig.~\ref{fig:scatter}b). Person correlation coefficients $r(H,E)=0.57$, coefficient of determination $R^2=0.3234$, shows moderate positive linear correlation and  $r(H,S)=0.9518$, $R^2=0.9059$, shows strong positive linear correlation. While Shannon entropy $H$ vs space filling $D$ (Fig.~\ref{fig:scatter}c) show very weak negative correlations, $r(H,D)=-0.1722$, $R^2=0.0297$.

Based on measures calculated we can construct the following hierarchies of complexity:
\begin{itemize}
\item CA representing 2-ary functions. 
\begin{itemize}
\item $LZ$: $\{f_8, f_7\} \gg \{f_1, f_3, f_4\} > f_{11} > f_2$ 
\item $H$: $\{f_8,f_7\} \gg f_1 > f_2 \gg \{f_{11}, f_3, f_4\}$
\item $S$: $\{f_8, f_7\} > f_1 \gg \{f_3, f_4\} > \{f_2, f_{11}\}$
\item $E$: $f_7 > f_8 \gg \{f_1, f_4, f_3\} > \{f_2, f_{11}\}$    
\end{itemize}
\item CA representing 4-ary functions. 
\begin{itemize}
\item $LZ$: $f_{20} > f_{19} > \{f_{12}, f_{14} \} > f_{40} $
\item $H$:  $f_{12} > \{f_{14}, f_{20}\} > \{f_{19}, f_{40} \}$
\item $S$:  $\{f_{12}, f_{20}\} > f_{14} > \{f_{19}, f_{40} \}$
\item $E$:  $ f_{20} > \{f_{12}, f_{14}\} > \{f_{19}, f_{40}\}$ 
\end{itemize}
\item CA representing 8-ary functions. 
\begin{itemize}
\item $LZ$: $f_{21} > f_{36} > f_{37}$
\item $H$:  $f_{21} > \{f_{37}, f_{36}\}$
\item $S$:  $f_{21} > f_{37} > f_{36}$
\item $E$:  $f_{37} \gg f_{21} > f_{36}$ 
\end{itemize}
\end{itemize}

For CA governed by 2-ary functions, $LZ$ hierarchy shows that functions $f_8$ and $f_7$ have the highest complexity, significantly higher than the others, $f_1$,$f_3$, and $f_4$ have moderate complexity, $f_{11}$ lower, and $f_2$ the lowest. In Shannon complexity hierarchy $f_8$ and $f_7$ again rank highest, $f_1$ is slightly lower, followed by $f_2$; functions $f_{11}$,$f_3$, and $f_4$ rank lowest and are grouped together. Simpson index ordering indicates that $f_8$ and $f_7$ have the highest structural complexity, $f_1$ follows, with $f_3$ and $f_4$ significantly lower, and $f_2$ and $f_{11}$ the lowest. Order of expressive complexity puts function $f_7$ as  the highest, slightly higher than $f_8$;  functions $f_1$,$f_4$, and $f_3$ are moderate, while $f_2$ and $f_{11}$ rank the lowest.

For CA governed by 4-ary functions, the order of compressibility demonstrates that function $f_{20}$ has the highest complexity, followed by $f_{19}$, functions $f_{12}$ and $f_{14}$ have moderate complexity, and $f_{40}$ the lowest. Shannon complexity demonstrates that function $f_{12}$ ranks highest, with $f_{14}$ and $f_{20}$ following, $f_{19}$ and $f_{40}$ are the lowest. In Simpson hierarchy functions $f_{12}$ and $f_{20}$ are highest, followed by $f_{14}$, and $f_{19}$ and $f_{40}$ rank the lowest. In the expressiveness hierarchy function $f_{20}$ is highest, with $f_{12}$ and $f_{14}$ in the middle, and $f_{19}$ and $f_{40}$ the lowest.

In CA governed by 8-ary functions, compressibility hierarchy is the following. Function $f_{21}$ has the highest complexity, followed by $f_{36}$f and $f_{37}$. In the Shannon entropy and Simpson index hierarchies function $f_{21}$ is highest, with $f_{37}$ and $f_{36}$ being equal and lower. Expressiveness hierarchy shows that function $f_{37}$ is significantly higher, followed by $f_{21}$, and $f_{36}$ the lowest.

Functions $f_8$ and $f_7$ are consistently ranked highest across multiple criteria for 2-ary functions, indicating their higher complexity or influence. Function $f_{21}$ is ranked highest in the majority of criteria for 8-ary functions. Different criteria ($LZ$, $H$, $S$, $E$) can yield different hierarchies. For instance, in 4-ary functions, $f_{12}$ is ranked highest by $H$ and $S$ but not by $LZ$ or $E$. Expressiveness measure $E$ seems to have distinct rankings compared to others, especially in the 8-ary functions.

Functions which produce CA patterns with absolute highest Liv-Zempel complexity, Shannon entropy and Simpson diversity are $f_{21}$ (Tab.~\ref{tab:functionszno}c and Fig.~\ref{fig:8neighbourhod}a), $f_8$  (Tab.~\ref{tab:functionszno}a and 
Fig.~\ref{fig:binaryneighbourhod}f) and $f_7$ (Tab.~\ref{tab:functionszno}a and 
Fig.~\ref{fig:binaryneighbourhod}e). A function with highest expressiveness is $f_{37}$ (Tab.~\ref{tab:sop_2bit_prot}c and Fig.~\ref{fig:8neighbourhod}c). Whilst space-time configurations of CA governed by $f_37$ shows complex local dynamics, the global dynamics is dull. This shows that the expressiveness might be not a reliable measure of global complexity. If we normalise values of $LZ$, $H$ and $S$ complexity measures by a number of terms or literals, we will fine that functions $f_7$ and $f_8$ are most complex functions, relative to formula complexity and in terms of space-time dynamics, discovered in colloids.

\section{Conclusion and Discussion}

The colloid automata --- one-dimensional cellular automata (CA) governed by Boolean functions derived from ZnO, proteinoid, and their mixture, colloids --- exhibit a rich spectrum of space-time evolution. Using complexity measures such as Lempel-Ziv complexity, Shannon entropy, Simpson diversity, and expressiveness, we can construct families of complexity hierarchies based on the space-time configurations of these colloid CA. These hierarchies reflect the inherent complexities of the Boolean functions and provide a means to compare and understand their behaviour across different dimensions. 

\begin{figure}[!tbp]
    \centering
   \subfigure[]{\includegraphics[width=0.49\textwidth]{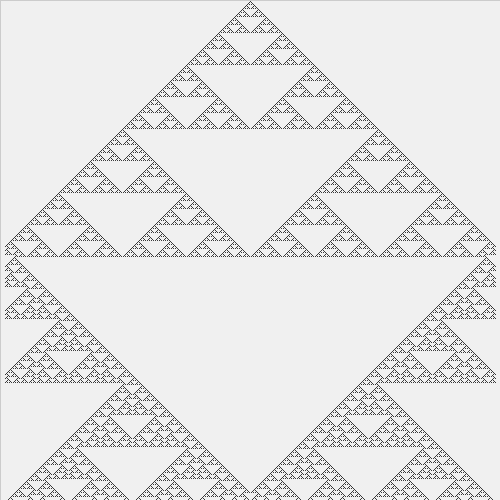}}
   \subfigure[]{\includegraphics[width=0.49\textwidth]{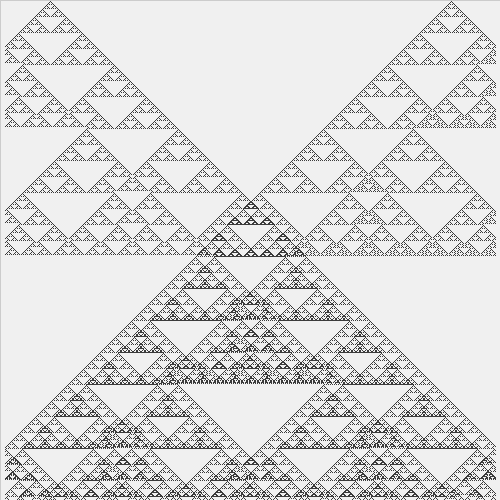}}
    \caption{Space-time evolution of one-dimensional CA governed by $f_7$. (a)~Initial configuration is a single cell (in the middle of the array) being in the state `1', all others are `0'. 
    (a)~Two cells, based 200 cells to the left and to the right the array's centre are at state `1' at the beginning of evolution. Cells in state `1' are black, `0' are light-gray. Time goes top down.}
    \label{fig:Sierpinski}
\end{figure}

The most complex, in terms of CA dynamics, functions discovered are {\sc xor} (function $f_7$) and {\sc not xor} (function $f_8$). The {\sc xor} gate is the most hard to find in natural non-linear systems, Boolean gate~\cite{boyar2000multiplicative,adamatzky2009complex}. The use of {\sc xor} gates in modern circuit design offers several advantages, such as reduced representation size and improved testability, and optimal power consumption~\cite{ye1999power}. CA governed by {\sc xor} gate exhibit unpredictable dynamics, similar to that of that randomly generated patterns \cite{martinez2024patterns} and, when evolve from single non-zero state configurations produced fractal patterns -- Sierpenski gasket~\cite{sierpinski1915courbe}. An evolution of rule $f_7$ CA started from a single cell in state `1' is shown in (Fig.~\ref{fig:Sierpinski}a), a reflection from absorbing boundaries is seen. The same evolution, but from two cells in state `1' (Fig.~\ref{fig
}a), shows a new fractal derived from a collision. The newly formed fractal pattern has a higher density of non-quiescent cells than the parent fractal structures.

There are several limitations of the research which could be addressed in future studies.  The research focuses solely on one-dimensional cellular automata. Extending this to two-dimensional or three-dimensional models could provide a more comprehensive understanding of the behaviour of colloid automata. The Boolean functions derived from ZnO, proteinoids, and their mixtures may not fully capture the complexities of information processing in real colloid systems. More sophisticated approaches incorporating physical and chemical interactions could yield more accurate results. The study is constrained by a finite set of states and rules, which might not encompass all possible behaviours of colloid systems. Exploring larger or infinite state spaces could reveal more complex dynamics.
The reliance on specific complexity measures such as Lempel-Ziv complexity, Shannon entropy, Simpson diversity, and expressiveness might not capture all aspects of the system's behaviour. Other measures or a combination of multiple metrics could provide a more holistic view. Measures like fractal dimension, Lyapunov exponents, or network-based metrics might offer new insights. Extending the research to higher-dimensional cellular automata could provide deeper insights into the spatial-temporal patterns of information processing in colloid systems, potentially revealing new patterns and behaviours.

\section*{Conflicts of interest}
There are no conflicts of interest to declare.

\section*{Availability of data}
The data are available on request.

\section*{Acknowledgements}
This project has received funding from the European Innovation Council And SMEs Executive Agency (EISMEA) under grant agreement No. 964388 ``COgITOR: A new colloidal cybernetic system towards 2030''.


\begin{thebibliography}{10}

\bibitem{adamatzky2019brief}
Andrew Adamatzky.
\newblock A brief history of liquid computers.
\newblock {\em Philosophical Transactions of the Royal Society B},
  374(1774):20180372, 2019.

\bibitem{chiolerio2017smart}
Alessandro Chiolerio and Marco~B Quadrelli.
\newblock Smart fluid systems: the advent of autonomous liquid robotics.
\newblock {\em Advanced Science}, 4(7):1700036, 2017.

\bibitem{chiolerio2020liquid}
Alessandro Chiolerio.
\newblock Liquid cybernetic systems: the fourth-order cybernetics.
\newblock {\em Advanced Intelligent Systems}, 2(12):2000120, 2020.

\bibitem{kheirabadi2023learning}
Noushin~Raeisi Kheirabadi, Alessandro Chiolerio, Neil Phillips, and Andrew
  Adamatzky.
\newblock Learning in colloids: Synapse-like zno+ dmso colloid.
\newblock {\em Neurocomputing}, 557:126710, 2023.

\bibitem{crepaldi2023experimental}
Marco Crepaldi, Charanraj Mohan, Erik Garofalo, Andrew Adamatzky, Konrad
  Szaci{\l}owski, and Alessandro Chiolerio.
\newblock Experimental demonstration of in-memory computing in a ferrofluid
  system.
\newblock {\em Advanced Materials}, 35(23):2211406, 2023.

\bibitem{roberts2023logical}
Nic Roberts, Noushin~Raeisi Kheirabadi, Michail-Antisthenis Tsompanas,
  Alessandro Chiolerio, Marco Crepaldi, and Andrew Adamatzky.
\newblock Logical circuits in colloids.
\newblock {\em arXiv preprint arXiv:2307.02664}, 2023.

\bibitem{fortulan2023reservoir}
Raphael Fortulan, Noushin~Raeisi Kheirabadi, Panagiotis Mougkogiannis,
  Alessandro Chiolerio, and Andrew Adamatzky.
\newblock Reservoir computing with colloidal mixtures of zno and proteinoids.
\newblock {\em arXiv preprint arXiv:2312.08130}, 2023.

\bibitem{verstraeten2007experimental}
David Verstraeten, Benjamin Schrauwen, Michiel d’Haene, and Dirk Stroobandt.
\newblock An experimental unification of reservoir computing methods.
\newblock {\em Neural networks}, 20(3):391--403, 2007.

\bibitem{lukovsevivcius2009reservoir}
Mantas Luko{\v{s}}evi{\v{c}}ius and Herbert Jaeger.
\newblock Reservoir computing approaches to recurrent neural network training.
\newblock {\em Computer Science Review}, 3(3):127--149, 2009.

\bibitem{dale2017reservoir}
Matthew Dale, Julian~F Miller, and Susan Stepney.
\newblock Reservoir computing as a model for in-materio computing.
\newblock In {\em Advances in Unconventional Computing}, pages 533--571.
  Springer, 2017.

\bibitem{konkoli2018reservoir}
Zoran Konkoli, Stefano Nichele, Matthew Dale, and Susan Stepney.
\newblock Reservoir computing with computational matter.
\newblock In {\em Computational Matter}, pages 269--293. Springer, 2018.

\bibitem{dale2019substrate}
Matthew Dale, Julian~F Miller, Susan Stepney, and Martin~A Trefzer.
\newblock A substrate-independent framework to characterize reservoir
  computers.
\newblock {\em Proceedings of the Royal Society A}, 475(2226):20180723, 2019.

\bibitem{miller2002evolution}
Julian~F Miller and Keith Downing.
\newblock Evolution in materio: Looking beyond the silicon box.
\newblock In {\em Proceedings 2002 NASA/DoD Conference on Evolvable Hardware},
  pages 167--176. IEEE, 2002.

\bibitem{miller2014evolution}
Julian~F Miller, Simon~L Harding, and Gunnar Tufte.
\newblock Evolution-in-materio: evolving computation in materials.
\newblock {\em Evolutionary Intelligence}, 7(1):49--67, 2014.

\bibitem{stepney2019co}
Susan Stepney.
\newblock Co-designing the computational model and the computing substrate.
\newblock In {\em International Conference on Unconventional Computation and
  Natural Computation}, pages 5--14. Springer, 2019.

\bibitem{miller2018materio}
Julian~F Miller, Simon~J Hickinbotham, and Martyn Amos.
\newblock In materio computation using carbon nanotubes.
\newblock In {\em Computational Matter}, pages 33--43. Springer, 2018.

\bibitem{miller2019alchemy}
Julian~Francis Miller.
\newblock The alchemy of computation: designing with the unknown.
\newblock {\em Natural Computing}, 18(3):515--526, 2019.

\bibitem{roberts2023mining}
Nic Roberts and Andrew Adamatzky.
\newblock Mining logical circuits in fungi.
\newblock In {\em Fungal Machines: Sensing and Computing with Fungi}, pages
  311--321. Springer, 2023.

\bibitem{shannon1948mathematical}
Claude~Elwood Shannon.
\newblock A mathematical theory of communication.
\newblock {\em The Bell system technical journal}, 27(3):379--423, 1948.

\bibitem{lin1991divergence}
Jianhua Lin.
\newblock Divergence measures based on the shannon entropy.
\newblock {\em IEEE Transactions on Information theory}, 37(1):145--151, 1991.

\bibitem{eskov2017shannon}
VM~Eskov, VV~Eskov, Yu~V Vochmina, DV~Gorbunov, and LK~Ilyashenko.
\newblock Shannon entropy in the research on stationary regimes and the
  evolution of complexity.
\newblock {\em Moscow university physics bulletin}, 72:309--317, 2017.

\bibitem{somerfield2008simpson}
PJ~Somerfield, KR~Clarke, and RM~Warwick.
\newblock Simpson index.
\newblock In {\em Encyclopedia of ecology}, pages 3252--3255. Elsevier, 2008.

\bibitem{nagendra2002opposite}
Harini Nagendra.
\newblock Opposite trends in response for the shannon and simpson indices of
  landscape diversity.
\newblock {\em Applied geography}, 22(2):175--186, 2002.

\bibitem{kim2017deciphering}
Bo-Ra Kim, Jiwon Shin, Robin~B Guevarra, Jun~Hyung Lee, Doo~Wan Kim, Kuk-Hwan
  Seol, Ju-Hoon Lee, Hyeun~Bum Kim, and Richard~E Isaacson.
\newblock Deciphering diversity indices for a better understanding of microbial
  communities.
\newblock {\em Journal of Microbiology and Biotechnology}, 27(12):2089--2093,
  2017.

\bibitem{ziv1977universal}
Jacob Ziv and Abraham Lempel.
\newblock A universal algorithm for sequential data compression.
\newblock {\em IEEE Transactions on information theory}, 23(3):337--343, 1977.

\bibitem{adamatzky2012phenomenology}
Andrew Adamatzky and Leon~O Chua.
\newblock Phenomenology of retained refractoriness: On semi-memristive discrete
  media.
\newblock {\em International Journal of Bifurcation and Chaos}, 22(11):1230036,
  2012.

\bibitem{redeker2013expressiveness}
Markus Redeker, Andrew Adamatzky, and Genaro~J Mart{\`\i}nez.
\newblock Expressiveness of elementary cellular automata.
\newblock {\em International Journal of Modern Physics C}, 24(03):1350010,
  2013.

\bibitem{adamatzky2018generative}
Andrew Adamatzky.
\newblock Generative complexity of gray--scott model.
\newblock {\em Communications in Nonlinear Science and Numerical Simulation},
  56:457--466, 2018.

\bibitem{roelofs1999png}
Greg Roelofs.
\newblock {\em PNG: the definitive guide}.
\newblock O'Reilly \& Associates, Inc., 1999.

\bibitem{howard1993design}
Paul~Glor Howard.
\newblock {\em The design and analysis of efficient lossless data compression
  systems}.
\newblock Brown University, 1993.

\bibitem{deutsch1996zlib}
Peter Deutsch and Jean-Loup Gailly.
\newblock Zlib compressed data format specification version 3.3.
\newblock Technical report, 1996.

\bibitem{wolfram1983statistical}
Stephen Wolfram.
\newblock Statistical mechanics of cellular automata.
\newblock {\em Reviews of modern physics}, 55(3):601, 1983.

\bibitem{boyar2000multiplicative}
Joan Boyar, Ren{\'e} Peralta, and Denis Pochuev.
\newblock On the multiplicative complexity of boolean functions over the basis
  (∧,⊕, 1).
\newblock {\em Theoretical Computer Science}, 235(1):43--57, 2000.

\bibitem{adamatzky2009complex}
Andy Adamatzky and Larry Bull.
\newblock Are complex systems hard to evolve?
\newblock {\em Complexity}, 14(6):15--20, 2009.

\bibitem{ye1999power}
Yibin Ye, Kaushik Roy, and Rolf Drechsler.
\newblock Power consumption in xor-based circuits.
\newblock In {\em Proceedings of the ASP-DAC'99 Asia and South Pacific Design
  Automation Conference 1999 (Cat. No. 99EX198)}, pages 299--302. IEEE, 1999.

\bibitem{martinez2024patterns}
Genaro~J Mart{\'\i}nez, Andrew Adamatzky, Rolf Hoffmann, Dominique
  D{\'e}s{\'e}rable, and Ivan Zelinka.
\newblock Patterns and dynamics of rule 22 cellular automaton.
\newblock In {\em ACTIN COMPUTATION: Unlocking the Potential of Actin Filaments
  for Revolutionary Computing Systems}, pages 37--86. World Scientific, 2024.

\bibitem{sierpinski1915courbe}
Warclaw Sierpinski.
\newblock Sur une courbe dont tout point est un point de ramification.
\newblock {\em CR Acad. Sci.}, 160:302--305, 1915.

\end{thebibliography}
\end{document}